\documentclass[english,twocolumn,aps,prl,showpacs]{revtex4-1}

\usepackage[T1]{fontenc}
\usepackage[latin1]{inputenc}
\usepackage{graphicx}
\usepackage{amssymb}
\usepackage{amsfonts}
\usepackage{amsmath}
\usepackage{epsfig}

\makeatletter

\input{epsf}

\makeatother

\usepackage{babel}
\makeatother

\begin{document}

\title{Temperature dependence of the contact in a unitary Fermi gas}

\author{E. D. Kuhnle, S. Hoinka, P. Dyke, H. Hu, P. Hannaford, and C. J. Vale}
\affiliation{\ ARC Centre of Excellence for Quantum-Atom Optics, Centre for Atom Optics and Ultrafast Spectroscopy, \\
Swinburne University of Technology, Melbourne 3122, Australia}

\date{\today}

\begin{abstract}
The contact ${\cal I}$, introduced by Tan, has emerged as a key parameter characterizing universal properties of strongly interacting Fermi gases.  For ultracold Fermi gases near a Feshbach resonance, the contact depends upon two quantities: the interaction parameter $1/(k_F a)$, where $k_F$ is the Fermi wave-vector and $a$ is the $s$-wave scattering length, and the temperature $T/T_F$, where $T_F$ is the Fermi temperature.  We present the first measurements of the temperature dependence of the contact in a unitary Fermi gas using Bragg spectroscopy.  The contact is seen to follow the predicted decay with temperature and shows how pair-correlations at high momentum persist well above the superfluid transition temperature.

\end{abstract}


\maketitle

Two component ultracold Fermi gases near Feshbach resonances provide an archetypal setting to explore universal behaviors \cite{heiselberg,ho,hunatphys}.  Universal systems should satisfy two requirements: firstly, the gas must be dilute enough that the mean interparticle spacing $n^{-1/3}$ greatly exceeds the range of the interaction potential $r_0$, and, secondly, the interactions, characterized by the $s$-wave scattering length, $a$, should be sufficiently strong that $a$ greatly exceeds $n^{-1/3}$.  All Fermi systems that satisfy these requirements will behave identically on a scale given by the average particle separation, independent of the details of the interaction potential.  Recent theoretical work by Tan \cite{tan1,tan2,tan3} and others \cite{werner,zhang,braaten,combescot2,huepl} has identified several exact relations applicable to Fermi systems in the universal regime.  The central parameter in these relations is a quantity called the contact ${\cal I}$, which forms a link between microscopic and macroscopic system properties.  

Contact quantifies the likelihood of finding two interacting fermions with very small separation and is closely linked to the pair-correlation function \cite{tan1}.  In strong analogy with the phase diagram of the Bose-Einstein condensate (BEC) to Bardeen-Cooper-Schrieffer (BCS) superfluid crossover of two-component Fermi gases \cite{perali}, $\cal{I}$ depends upon two parameters: the dimensionless interaction strength $1/(k_F a)$, where $k_F$ is the Fermi wave-vector, and the relative temperature $T/T_F$, where $T_F$ is the Fermi temperature.  Previous theoretical \cite{werner,palestini} and experimental \cite{partridge,stewartgaebler,kuhnle} work has investigated the interaction dependence of the contact, and a number of recent studies have calculated the temperature dependence of the contact \cite{yu,palestini,enss,huarxiv}.  To date however, there have been no measurements of this latter dependence.

In this letter, we report the first measurements of the temperature dependence of the contact using Bragg spectroscopy of a $^6$Li Fermi gas at unitarity.  Bragg spectroscopy allows for quantitative measurements of the static structure factor $S(k)$ which is directly proportional to the contact at high momenta.  We extract the first and second moments from our Bragg spectra and use these to obtain the dynamic structure factor $S(k,\omega)$ and from this $S(k)$ and the contact.  Our results are in good agreement with theoretical predictions and indicate that pair-correlations at high momenta persist well above the critical temperature for superfluidity.

Tan's exact relations for Fermi gases near the BEC-BCS crossover marked a dramatic development in the understanding of highly correlated Fermi systems \cite{tan1,tan2,tan3}.  Strongly interacting Fermi gases represent a difficult theoretical challenge as the large scattering length leaves no small parameter that can be used in perturbative theories.  Having invoked the contact, Tan was able to derive a number of relations linking the microscopic parameters to macroscopic properties such as the energy and momentum distribution.  Contact is defined as ${\cal I} \equiv \lim_{k \rightarrow  \infty} k^4 n_\sigma(k)$ where the momentum density of a particular spin component $n_\sigma(k)$ decays with $1/k^4$ at large $k$.  Thus $\cal{I}$ quantifies the amplitude of this high momentum tail.  A more intuitive understanding of ${\cal I}$ is evident through its relation to the two-body correlation function between spin-up $(\uparrow)$ and spin-down $(\downarrow)$ fermions
\begin{equation}
g_{\uparrow \downarrow}^{(2)}(r) = \frac{{\cal I}}{16\pi ^2} \left(\frac{1}{r^2}-\frac{2}{ar}\right),
\end{equation}
valid for $r_0 < r < k_F^{-1}$, where the contact appears as a prefactor \cite{tan1}.  The contact therefore quantifies the likelihood of finding two fermions at distances small compared to the many-body length scales.

Contact is closely linked to the pairing temperature $T^*$ in the high momentum limit. A number of experiments have measured the contact using photo-association \cite{partridge,werner}, radio-frequency spectrocopy \cite{stewartgaebler}, the tail of the momentum distribution \cite{stewartgaebler} and Bragg spectroscopy \cite{kuhnle}.  These measurements demonstrated the predicted decay of the contact through the transition from the BEC to BCS sides of a Feshbach resonance.  Pairing ($T^*$) in the phase diagram of the BEC-BCS crossover is also strongly temperature dependent \cite{perali}.  Our measurements demonstrate the build up of pair-correlations from $T \gg T^*$ down to $T \rightarrow 0$, in the momentum range where Tan's relations hold.

The Fourier transform of $g_{\uparrow \downarrow}^{(2)}(r)$ yields the static structure factor $S_{\uparrow\downarrow}(k)$ \cite{pitaevskii,combescot}, which can be quantitatively measured using Bragg spectroscopy \cite{kuhnle}.  In the limit where $a \rightarrow \infty$ the second term in Eq. 1 vanishes and the $S_{\uparrow\downarrow}(k)$ is given by
\begin{equation}
S_{\uparrow\downarrow}\left(k \gg k_F\right) = \left ( \frac{\cal I}{N k_F} \right ) \frac{k_F}{4 k}, 
\end{equation}
where ${\cal I}/(N k_F)$ is the dimensionless contact and $N$ is the total number of atoms. At large momenta, $k \gg k_F$, the total static structure factor $S(k) \cong 1 + S_{\uparrow\downarrow}(k)$ as the spin-parallel component approaches the uncorrelated value of unity at large $k$ \cite{combescot,veeravalli}.  Therefore we can readily determine the spin-antiparallel component, $S_{\uparrow\downarrow}(k)$, by measuring the total structure factor.

Experimentally we create a unitary Fermi gas by evaporatively cooling a balanced mixture of $^6$Li atoms in the two lowest spin states $\vert F$ = 1/2, $m_F$ = $\pm$ 1/2 $\rangle$ at a magnetic field of 834 G in an optical dipole trap.  After transferring the atoms into a second deep dipole trap, we obtain $N_\sigma = 170,000$ atoms in each spin state at a temperature of 0.09 $\pm$ 0.03 $T/T_F$. The trap frequencies in the final trap are $(\omega_{x},\omega_{y},\omega_{z})/2 \pi = (24.5,65,230)$ Hz, ($\bar{\omega}$ = 71.5 Hz) giving a Fermi energy $E_F/(2 \pi \hbar)$ = 7.2~kHz, where $E_F= \hbar\bar{\omega} \left(3 N \right)^{1/3}$ and $k_F = \sqrt{2 m E_F}/\hbar$ = 2.9 $\mu m^{-1}$.  At 834 G the $s$-wave scattering length diverges and collisions are unitarity limited.

We vary the temperature of the cloud by suddenly switching off the dipole trap for a variable time before quickly ramping it on and holding the cloud for 400 ms ($\gg 1 / \bar{\omega}$) to rethermalize \cite{kinast}.  In this way we can repeatably heat the cloud to temperatures up to 1.1 $T/T_F$ without loss of atoms.  To determine the temperature, we image clouds at unitarity after 2 ms expansion and obtain an empirical temperature $\tilde{T}$ by fitting Thomas-Fermi profiles to the images \cite{kinast}.  While the conversion from $\tilde{T}$ to the true temperature $T/T_F = \tilde{T} \sqrt{1+\beta}$, where $\beta \approx -0.58$ is the universal parameter \cite{giorgini}, is not exact, we have independently calibrated it through measurements of the mean energy per particle and compared these to predictions based on a Nozi\`{e}re-Schmitt-Rink (NSR) theory \cite{hupra73} which agrees well with thermodynamic measurements of the equation of state \cite{nascimbene1,horikoshi,hunjp}. The two methods agree to better than 0.05 $T_F$ over the full range of temperatures studied here.

Bragg scattering is achieved by illuminating a cloud of atoms with two laser beams that have a small frequency difference $\omega$ and intersect at an angle $\theta$.  This creates a periodic potential which moves at a velocity $\omega/k$, where $k =4 \pi \sin{(\theta/2)}/\lambda$ and $\lambda$ is the wavelength of the Bragg light.  Measuring the response of the cloud to a sequence of Bragg pulses as a function of $\omega$, yields a Bragg spectrum from which quantitative information on the dynamic structure factor can be obtained \cite{brunello}. 

To perform Bragg spectroscopy, we use two laser beams that intersect at $\theta = 49.5^{\circ}$.  At this angle, $k = 2.7 k_F$ and the resonant recoil frequency for Bragg scattering of free atoms is $\omega_r / (2 \pi)$ = 51.6 kHz.  The Bragg lasers are detuned $\sim$2.5 GHz from the scattering transition to avoid spontaneous scattering and the Bragg pulse duration is 200 $\mu$s.  For our beam intensities this duration is well below the two-photon Rabi cycling period ensuring spectra are obtained in the linear response regime.  Each of the two spin states in the $^6$Li gas couple almost equally to the Bragg lasers to within $4\%$.  We measure the resulting atomic distribution, $n(x,y)$, a further 3 ms after applying the Bragg pulse.  The trap laser is switched off immediately after the Bragg pulse.  Both Bragg scattering and imaging take place at 834 G.  From these images we obtain line profiles $n(x)$ by integrating over the $y$-direction perpendicular to the Bragg scattering.

We perform a series of experiments to acquire a sequence of profiles $n(x)$ as the frequency difference between the two Bragg lasers $\omega$ is varied.  Quantitative analysis is achieved by evaluating the first and second moments $\langle x \rangle$ and $\langle x^2 \rangle$, respectively, for every line profile, where $\langle x^m \rangle = \sum_i x_i^m n(x_i)  / \sum_i n(x_i)$ and the sum is over all pixels $i$.  At large $\omega$ no Bragg scattering occurs and the moments $\langle x_0 \rangle$ and $\langle x_0^2 \rangle$ provide a reference center of mass position and mean square cloud width, respectively.  The difference between these reference moments and those obtained when Bragg scattering occurs provides two ways to quantify the effect of the Bragg pulse.  The center of mass displacement due to the Bragg pulse is given by $\Delta X(\omega) = \langle x(\omega) \rangle -  \langle x_0 \rangle$, which is proportional to the momentum transferred to the cloud, and the increase in mean square cloud size $\Delta \sigma_x^2(\omega) = \langle x^2(\omega) \rangle -  \langle x_0^2 \rangle$ is proportional to the energy transferred \cite{brunello}.  Spectra obtained the the first and second moments, at temperatures of $T/T_F$ = 0.10, 0.29, 0.58 and 1.0, are presented in Fig. 1(a) and (b), respectively.  Each spectrum is an average of four individual spectra.

\begin{figure}[htbp]
\centering
\includegraphics[clip=true,width=0.48\textwidth]{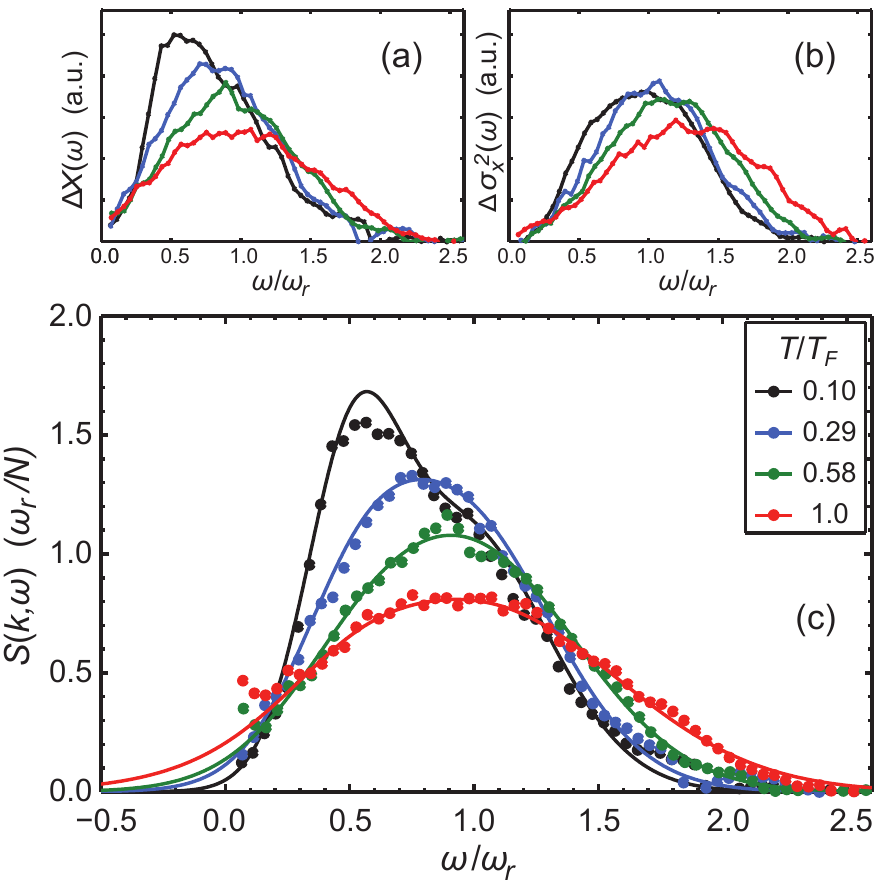}
\caption{Bragg spectra and dynamic structure factor of a unitary Fermi gas. Measured change in (a) the first moment (center of mass displacement) $\Delta X(\omega)$ and (b) the second moment (increase in mean square size) $\Delta \sigma_x^2(\omega)$ as a function of the Bragg frequency $\omega$ at temperatures $T/T_F$ = 0.10, 0.29, 0.58 and 1.0.  (c) Dynamic structure factor $S(k,\omega)$ obtained from a weighted average of the first and second moments.  Points are the experimental data and the solid lines are double Gaussian fits normalized according to the $f$-sum rule. }
\label{energy}
\end{figure}

As the duration of the Bragg pulse is relatively long, the effects of Fourier broadening on our spectra are small and we ignore them in our analysis.  For a given run of the experiment, with a particular value of $\omega$, the momentum $\{$energy$\}$ transferred to the cloud will be proportional to $k \cdot \left [S(k,\omega)-S(-k,-\omega) \right]$ $\{ \omega \cdot \left [S(k,\omega)-S(-k,-\omega) \right] \}$ \cite{brunello}.  Apart from the factors of $k$ and $\omega$, the pre-factors multiplying the difference of the positive and negative components of the dynamic structure factor are identical, so the energy transferred is equal to $\omega/k$ times the momentum.  We include the negative component $S(-k,-\omega)$ in our analysis as this describes de-excitation from high lying states which can become significant at the higher temperatures and momentum ($k = 2.7k_F$) we consider.  To account for this, the principle of detailed balancing is employed which states that $S(-k,-\omega) = \exp{(-\frac{\hbar \omega}{k_B T})} S(k,\omega)$ where $k_B$ is Boltzmann's constant.  Thus both the first and second moments provide a measure of $S(k,\omega)$, as $k, \omega$ and $T$ are all known.

In Fig.~1(c) we show the experimentally determined dynamic structure factors obtained from a weighted average of  $\Delta x (\omega) / k$ and $\Delta \sigma_x^2 (\omega) / \omega$, for the same temperatures.  The weighting of each contribution is inversely proportional to the relative uncertainty in the respective measurement points.  The detailed balance term [$1-\exp{(-\frac{\hbar \omega}{k_B T}})$] has also been divided out to yield $S(k,\omega)$.  This increases the noise in the data at low frequencies where this term becomes small, particularly at higher temperatures.  Along with the data points are fits based on two Gaussian functions, centered near $\omega_r/2$ and $\omega_r$ to account for pair and free atom excitations in $S(k,\omega)$, respectively.  While the true structure factors will not necessarily be well described by two Gaussians \cite{zou}, we have estimated that the errors introduced in approximating $S(k,\omega)$ with two Gaussians is at the level of a few percent for the coldest clouds and less at higher temperatures.  The spectra are normalized according to the $f$-sum rule \cite{pines}, such that the fitted functions for $S(k,\omega)$ satisfy the integral $\hbar \int \omega S(k,\omega) d\omega = N \omega_r$.  At low temperatures $S(k,\omega)$ is dominated by excitations at pair frequencies $\omega_r/2$ and the peak in $S(k,\omega)$ shifts towards $\omega_r$ as $T$ approaches $T_F$.  The change from the low $T$ to high $T$ limits occurs smoothly over the temperature range covered as expected from quantum virial expansion calculations \cite{hupra81}.  At high temperatures $S(k,\omega)$ contains significant weight at negative frequencies.

From these dynamic structure factors, we can obtain the static structure factor $S(k)$, defined by $N S(k) = \hbar \int S(k,\omega) d\omega$.  As we do not measure $S(k,\omega)$ at negative frequencies, and because of the enhanced noise in the experimental data at low frequencies, we use the integral of the double Gaussian fits to $S(k,\omega)$ over $\omega$ to obtain $S(k)$.  As these satisfy the $f$-sum rule the integral provides a robust measure of $S(k)$ \cite{kuhnle}.  Equation (2) allows us to directly link $S_{\uparrow \downarrow}(k) = S(k)-1$ to the contact.  Extracting the contact from the dynamic structure factors (Fig. 1(c)) for all measured temperatures gives the data shown in Fig. 2.  The vertical error bars are the statistical uncertainty based on the range of values obtained from the individual spectra and the horizontal error bars indicate the uncertainty in the temperature measurement.  Also shown in this figure are different theoretical calculations for $\cal{I}$.  The three full lines are obtained from strong-coupling theories based on the many-body $t$-matrix approximation: a non-self-consistent (G$_0$G$_0$) theory (purple) \cite{palestini}, a self-consistent (GG) theory (green) \cite{enss} and a Gaussian pair fluctuation theory (GPF) (brown) \cite{huarxiv}.  The dashed gray lines show the results of a quantum virial expansion to second (long dashed) and third (short dashed) order \cite{huarxiv}. 

Calculations of the zero temperature contact typically predict values between 3 and 3.4 \cite{combescot,palestini,enss,huarxiv} consistent with our measurement at the lowest temperature $3.11 \pm 0.23$ and an earlier value of $3.40 \pm 0.18$ extracted from measurements of the equation of state \cite{navon}.  At higher temperatures, the contact is seen to decay monotonically from $\sim 3$ down to below 0.5 over the range $0.1 - 1\, T_F$, in good general agreement with theory.  The different theoretical methods each predict slightly different behavior near the critical temperature for superfluidity $T_c \approx 0.2 \, T_F$.  The uncertainties in our measurements and limited number of data points prevent us from identifying one theory as more accurate than another.  With further improvements, and a closer study of the region near $T_c$, it should be possible to resolve the discrepancies between different theories.

\begin{figure}[htbp]
\begin{center}
\includegraphics[clip=true,width=0.47\textwidth]{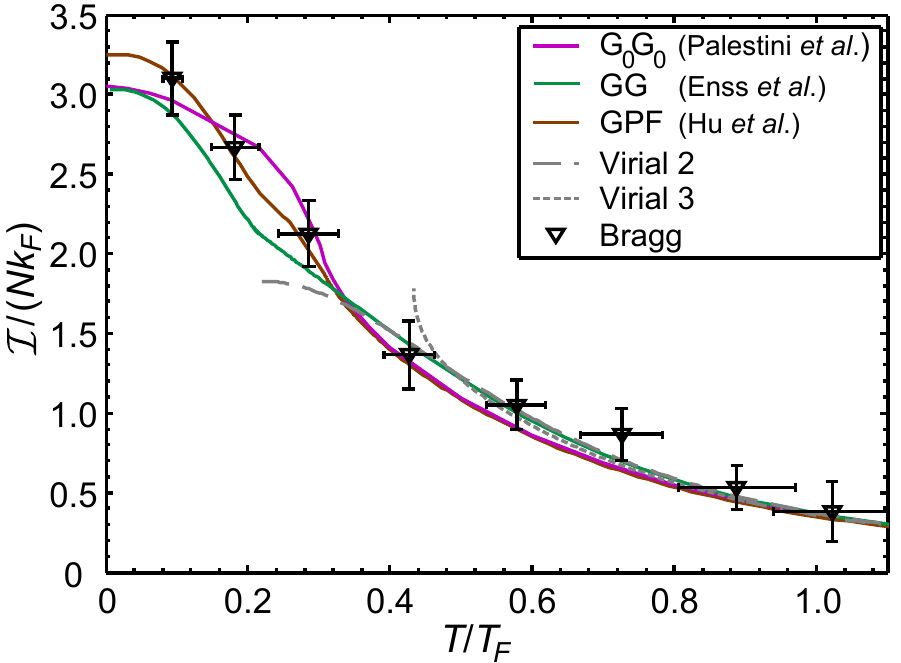}
\caption{Temperature dependence of the contact over the range $T/T_F = 0.1 - 1.0$. Data points are the measured values of $\cal{I}$ obtained from the static structure factors.  Solid lines are predictions based on three different strong-coupling theories as described in the text \cite{palestini,enss,huarxiv}.  The long-dashed and short-dashed lines are second and third order quantum virial expansion calculations, respectively \cite{huarxiv}.}
\label{Sqfinal}
\end{center}
\end{figure}

We note that $\cal{I}$ is significant at temperatures well above $T_c$ and shows the gradual build up of pair-correlations below $T_F$.  At the momentum we have studied ($2.7\hbar k_F$) the $1/k^4$ momentum tail dominates the pair-correlation function so these measurements should not be taken as evidence for pseudogap pairing.  Whether a connection between short-range ($r \ll k_F^{-1}$) and medium-range ($r \sim k_F^{-1}$) pair-correlations can be found remains an open, but important, question \cite{palestini} that may offer new insights into pairing both below \cite{stewart,schirotzek} and above $T_c$ \cite{gaebler}.  

In summary, we have presented the first measurements of the temperature dependence of Tan's contact in a strongly interacting Fermi gas.  These were achieved using Bragg spectroscopy to make quantitative measurements of the dynamic and static structure factors.  Our results indicate that the contact, and hence the high momentum component of the pair-correlation function, remains significant at temperatures well above $T_c$, in good agreement with theoretical predictions.  More extensive studies of the region near $T_c$ are needed to resolve the discrepancies between different theoretical approaches.  Extracting the homogeneous contact from experiments on trapped systems would allow a more sensitive probe of these discrepancies \cite{palestini,huarxiv}.  Bragg experiments at lower momentum ($k \sim k_F$) may also elucidate a connection between the contact and pseudogap pairing.

We thank R. Haussmann, T. Enss and F. Palestini for providing theoretical data and X.-J. Liu and P. Drummond for fruitful discussions.  This work is supported by the Australian Research Council Centre of Excellence for Quantum-Atom Optics.

\end{document}